\documentclass[pra,showpacs,twocolumn,aps]{revtex4}

\usepackage{color}

\usepackage{amsmath}
\usepackage{bm}
\usepackage{amssymb}

\usepackage{graphicx}
\usepackage{epsfig}
\usepackage{psfrag}

\begin{document}

\title{Quantum trajectory phase transitions in the micromaser}

\author{Juan P.  Garrahan}

\author{Andrew D. Armour}

\author{Igor Lesanovsky}

\affiliation{School of Physics and Astronomy, University of
Nottingham, Nottingham, NG7 2RD, UK}

\begin{abstract}
We study the dynamics of the single atom maser, or micromaser, by means of the recently introduced method of thermodynamics of quantum jump trajectories.  We find that the dynamics of the micromaser displays multiple {\em space-time} phase transitions, i.e., phase transitions in ensembles of quantum jump trajectories.  This rich dynamical phase structure becomes apparent when trajectories are classified by dynamical observables that quantify {\em dynamical activity}, such as the number of atoms that have changed state while traversing the cavity.  The space-time transitions can be either first-order or continuous, and are controlled not just by standard parameters of the micromaser but also by non-equilibrium ``counting'' fields.  We discuss how the dynamical phase behavior relates to the better known stationary state properties of the micromaser.  
\end{abstract}


\maketitle

\section{Introduction}

Dynamical transitions are a common feature of driven open quantum systems. Probably the best known example is the laser whose behavior close to threshold has many features in common with a thermodynamic system close to a continuous phase transition \cite{scully}. Another text-book example which has an especially rich dynamical phase behavior is the micromaser \cite{scully,haroche,Filipowicz1986,Englert2002,Walther2006} where a flux of atoms pass one at a time through a microwave cavity. As the flux of atoms is increased the cavity undergoes a continuous dynamical crossover, followed by a series of first-order like crossovers.  These crossovers are in practice rounded transitions that only become formally analogous to phase transitions in the limit where the number of atoms sent through the cavity during its lifetime becomes infinitely large \cite{Filipowicz1986,Englert2002,Walther2006}.  The dynamics of the micromaser has been investigated very successfully in experiment by measuring the state of the atoms emerging from the cavity \cite{Walther2006}, an approach which gives access to quantum trajectories of the system \cite{Plenio1998,cresser}.

In this work we investigate the connection between the dynamical transitions in the micromaser and thermodynamic phase transitions in a new way by analysing the statistical properties of quantum trajectories of the cavity field. We build on a recent Letter \cite{prl} where two of us introduced a thermodynamic formalism for the study of quantum jump trajectories \cite{Plenio1998} of open quantum systems \cite{Gardiner}.  This formalism \cite{prl}, based on the large-deviation (LD) method \cite{LD}, treats statistical ensembles of dynamical trajectories in a manner equivalent to the way in which equilibrium statistical mechanics treats ensembles of configurations.  This method allows the statistics of dynamical events to be understood using the conceptual framework of thermodynamics \cite{books}.  In the classical context, this ``statistical mechanics of trajectories'' approach has proved useful in uncovering {\em space-time} phase transitions (i.e. transitions in the space of temporal trajectories) in the dynamics of slowly relaxing systems such as glasses \cite{Garrahan}. In Ref.\ \cite{prl} we discussed how similar space-time transitions can be present in dissipative quantum systems, and showed as an example one such transition in the micromaser.  Here we go beyond this result by providing a comprehensive study of the dynamical phase structure of the micromaser using the LD approach.

\begin{figure}
\includegraphics[width=.9\columnwidth]{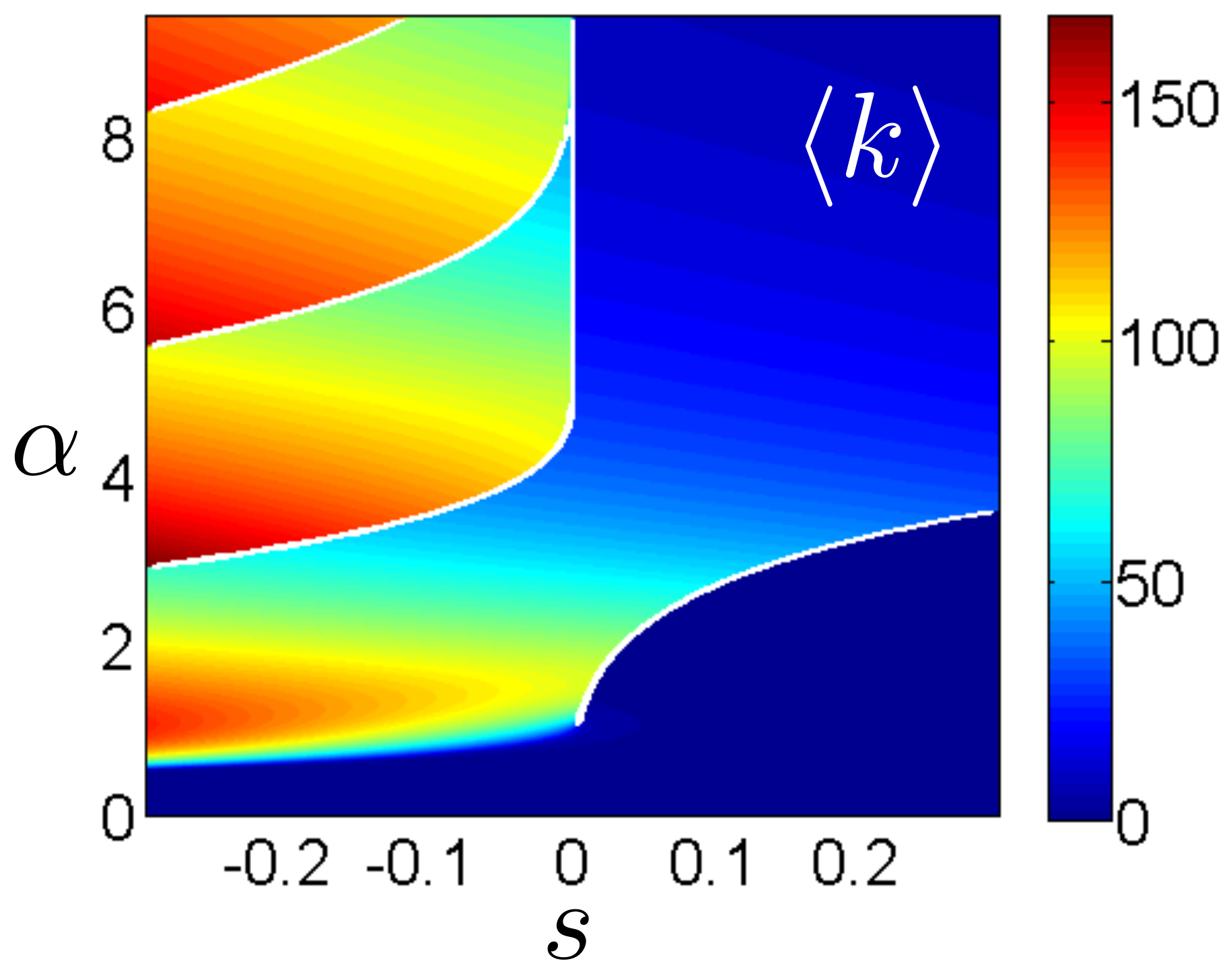}
\caption{
(Color online) Space-time phase diagram of the micromaser.
The dynamics displays multiple {\em space-time} transitions between dynamical phases of distinct dynamical {\em activity} $k$.  The phase diagram is in terms of the ``pump parameter'' $\alpha$, the standard control parameter of the micromaser, and the ``counting field'' $s$ (see text for details).   The white lines indicate first-order dynamic phase boundaries.  The rightmost phase boundary ends at a critical point with $s_c \gtrsim 0$ and $\alpha_c \gtrsim 1$.
}
\label{fig1}
\end{figure}

Figure \ref{fig1} summarizes the findings we describe in detail below.   It shows the ``space-time phase diagram" of the micromaser obtained via the method of \cite{prl}.  The figure is a density map for the average number of quantum jumps \cite{Plenio1998} corresponding to the average number of atoms, $\langle k\rangle$, which have released a quantum of energy into the cavity, per unit time and at stationary conditions.  This quantity is a {\em dynamical order parameter}, as it allows  distinct dynamical phases to be classified and distinguished between. The figure shows the dependence of $\langle k\rangle$ on two parameters, the reduced atom pump rate $\alpha$ (defined below) and the non-equilibrium ``counting'' field $s$ \cite{prl}. Whilst $\alpha$ can be varied directly in an experiment, the behavior at different values of the parameter $s$ describes the statistics of quantum jumps whose probability is biased by the value of $s$ so that they become more ($s<0$) or less ($s>0$) likely to occur. The behavior at $s=0$ describes the statistics of the trajectories which would be observed in an experiment. Remarkably, we find that at certain values of $\alpha$ and $s$ the order parameter changes in a way which appears to be {\it singular}, even though in this case only 100 atoms go through the cavity in its lifetime.  These points, which occur at non-zero values of $s$, are the location of phase transitions between phases of distinct ``dynamical activity'' \cite{prl,Garrahan,Lecomte,activity}.

The rest of this paper is organized as follows. In Sec.\ II we review the standard theoretical description of the micromaser and describe the quantum trajectories generated by measuring the state of the emerging atoms. We then introduce the counting field and describe how large deviation theory can be used to calculate the statistical properties of the trajectories. We use a mean-field approach to uncover the statistical properties of the trajectories for arbitrary $s$ in Sec.\ III, and hence obtain phase diagrams like that shown in Fig.\ \ref{fig1}. Then in Sec.\ IV we compare the mean-field results with an exact numerical analysis. Finally, Sec.\ V contains our conclusions and a short discussion.

\section{Model and formalism}

\begin{figure}
\includegraphics[width=.8\columnwidth]{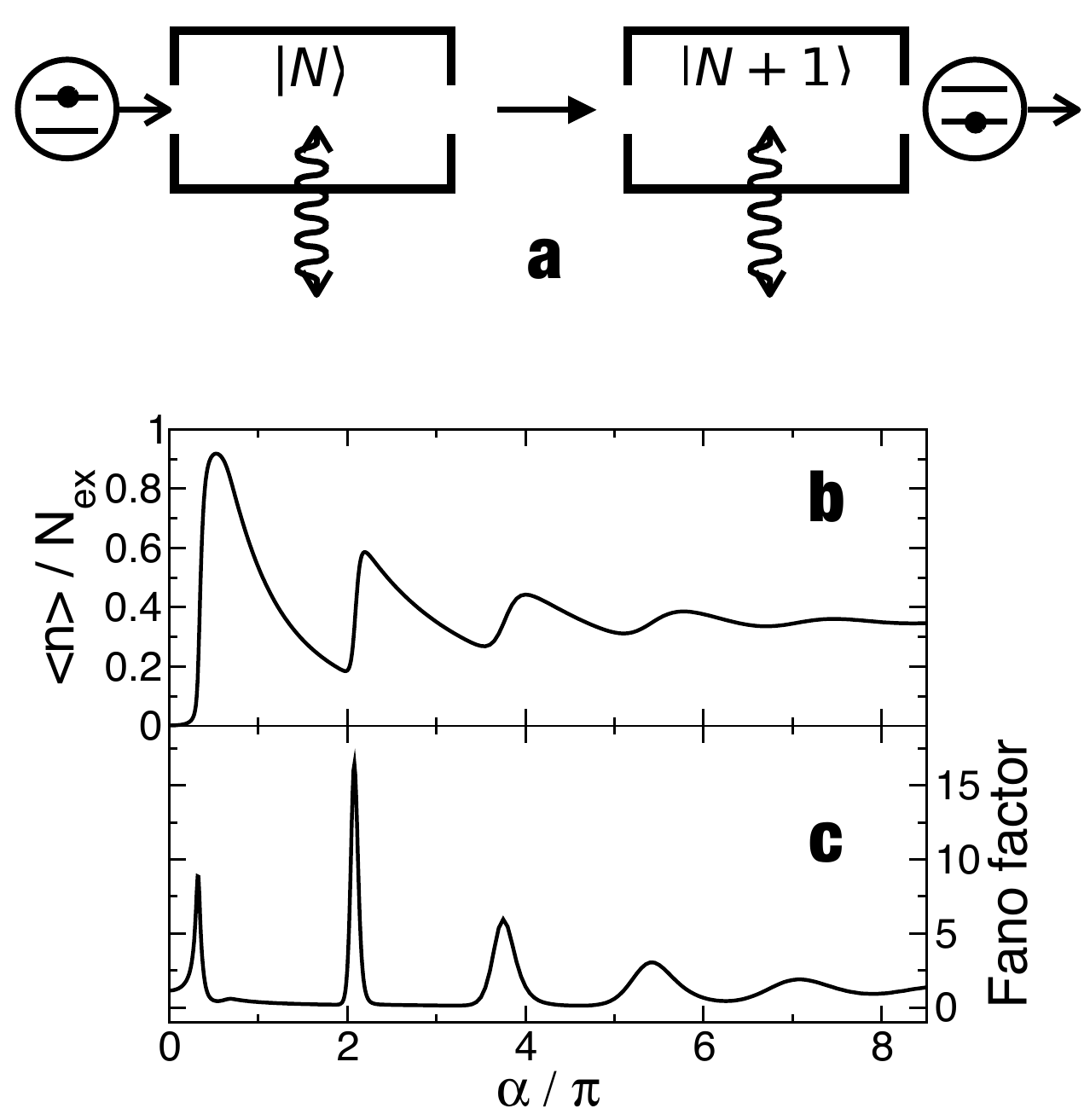}
\caption{
(a) Schematic operation of the micromaser \cite{Walther2006}.  The cavity mode interacts with a thermal bath, and is driven by two level atoms pumped, one by one, in their excited state.  When the atoms traverse the cavity they can decay to their ground state, releasing a photon that is absorbed by the cavity.  (b) Mean photon number $\langle n \rangle$ in the cavity in the steady state (in units of $N_{\rm ex}$) as a function of the ``pump parameter'', $\alpha \equiv \phi \sqrt{N_{\rm ex}}$.
Here $N_{\rm ex} = 100$ and $\nu = 0.15$.   (c) The Fano factor.}
\label{fig2}
\end{figure}

The micromaser consists of a microwave cavity which is pumped by a sequence of two level atoms which are sent into the cavity at a constant (pump) rate; the cavity is also coupled to a thermal bath \cite{Walther2006,Filipowicz1986,Englert2002,scully}. In the simplest case, which we will consider here, the atoms are all prepared in the excited state and pass through the cavity one at a time.  This setup is sketched in Fig.\ \ref{fig2}(a).  The atoms are assumed to interact resonantly with a single mode of the cavity.  The cavity mode reaches a steady state which is sensitive to the pump rate and the atom-cavity coupling as well as the properties of the thermal bath.

A standard treatment \cite{Filipowicz1986,Englert2002} for the micromaser focuses on the dynamics of the cavity.  After integrating out the thermal bath and the atom degrees of freedom, and under a Markovian approximation which assumes that the coupling of the cavity to both these systems is weak and that correlation times within these baths are much shorter than all relevant timescales, the density matrix of the cavity mode, $\rho$, evolves according to a quantum master equation (QME):
\begin{equation}
\dot{\rho} = {\cal W}(\rho) ,
\label{QME}
\end{equation}
where the super-operator ${\cal W}$ \cite{Englert2002} is of the Lindblad form \cite{Lindblad1976,Gardiner},
\begin{equation}
{\cal W}(\rho) = \sum_{\mu=1}^{4} \left( L_\mu \rho L_\mu^\dagger - \frac{1}{2} \{  L_\mu^\dagger L_\mu , \rho \} \right) ,
\label{W}
\end{equation}
and the Lindblad operators read,
\begin{eqnarray}
L_1 &=& \sqrt{N_{\rm ex}} ~ a^{\dagger}\frac{\sin \left( \phi \sqrt{aa^\dagger} \right)}{\sqrt{aa^\dagger}},
\label{L1} \\
L_2 &=& \sqrt{N_{\rm ex}} ~ \cos \left( \phi \sqrt{aa^\dagger} \right) ,
\label{L2} \\
L_3 &=& \sqrt{\nu +1} ~ a ,
\label{L3} \\
L_4 &=& \sqrt{\nu} ~ a^\dagger .
\label{L4}
\end{eqnarray}
Here $a,a^\dagger$ are the raising/lowering operators of the cavity mode, $N_{\rm ex}$
is the atom beam rate in units of the cavity lifetime (the inverse of the thermal relaxation rate) which we have set to one for simplicity, and $\nu$ is the average thermal photon occupation number.  The ``accumulated Rabi angle'' $\phi$ (the effective vacuum Rabi frequency of the atom-cavity system multiplied by the time spent by the atom in the cavity \cite{Englert2002})  quantifies the atom-cavity interaction. $L_1$ and $L_2$ are the quantum jump operators  on the cavity due to the interaction with the atoms describing absorption of a quantum from an atom and passage of an atom through the cavity without absorption respectively. The other two jump operators, $L_3$ and $L_4$,  are due to interaction with the thermal bath.

\subsection{Steady-state properties}
The stationary state density matrix of the QME (\ref{QME}-\ref{L4}) is easy to obtain exactly.  It is diagonal in the number basis, and reads \cite{Filipowicz1986,Englert2002}:\
\begin{equation}
\rho_{\rm s.s.}(n) = \rho_{\rm s.s.}(0) \prod_{m=1}^n \left( \frac{\nu}{\nu+1} + \frac{N_{\rm ex}}{\nu+1} \frac{\sin^2 \left(\phi \sqrt{m} \right)}{m} \right) .
\label{rhoss}
\end{equation}

An example of the steady-state behavior of the cavity as a function of the reduced atom pump parameter $\alpha=\phi\sqrt{N_{ex}}$ is shown in Figs. \ref{fig2}(b) and (c).
 As the pump parameter is increased from zero, the average number of cavity quanta, $\langle n \rangle$, displays a series of local maxima (minima)  corresponding to situations where the coupling between atom and cavity is such that the atom approximately performs a half (complete) Rabi cycle when traversing the cavity.  As the effective Rabi frequency depends on the occupation of the cavity, the sharpness of this effect is smoothed out.

 The regions of sudden increase in $\langle n\rangle$ that occur around $\alpha\simeq 1$ and $\alpha\simeq 2\pi,4\pi,...$ correspond to dynamical crossovers in the cavity state and are marked by sharp peaks in the corresponding Fano factor (defined as $(\langle n^2\rangle-\langle n\rangle^2)/\langle n\rangle$).  The crossovers at $\alpha\simeq 1$ is classified as continuous and the others as first-order based on an analysis of the behavior of $-\ln{[\rho_{\rm s.s.}(n)/\rho_{\rm s.s.}(0)]}$ in the spirit of Landau theory \cite{haroche} and on the observation that the regions of sudden increase in $\langle n\rangle$ around $\alpha\simeq 2\pi,4\pi,...$ (but not the one at $\alpha\simeq 1$) get progressively sharper as $N_{ex}$ is made larger.  Thus the crossovers are usually interpreted as being analogous to thermodynamic phase transitions in the limit $N_{ex}\rightarrow \infty$ \cite{Filipowicz1986,Walther2006}.  The number of minimas in $-\ln{[\rho_{\rm s.s.}(n)/\rho_{\rm s.s.}(0)]}$ increases progressively as $\alpha$ gets larger, but each individual minimum typically becomes less distinct which leads to the progressive broadening of the dynamical crossovers \cite{haroche}.

\subsection{Quantum Trajectories}
Instead of looking at the steady-state properties of the cavity density operator we will analyze the statistics of its quantum trajectories. 
The quantum jump trajectories of the system correspond to the time record of projection events due to the action of the Lindblad operators (\ref{L1}-\ref{L4}).  We wish to classify such trajectories according to the number of these events \cite{prl}.  The probability $P_t(K)$ to observe $K$ events after time $t$ is given by $P_t(K) = {\rm Tr} \left[ \rho^{(K)}(t) \right]$, where $\rho^{(K)}(t)$ is a reduced density matrix obtained by the projection of the full density matrix  (i.e. cavity and baths) onto the subspace of $K$ events \cite{Zoller} and $\rho(t)=\sum_K \rho^{(K)}(t)$.  Here we will consider in particular quantum jumps associated with the jump operator $L_1$ Eq.\ (\ref{L1}) which describes the detection of unexcited atoms emerging from the cavity, as this type of measurement is readily performed in experiment. However, in principle we could define a counting variable associated with any of the four quantum jump operators and the results would be equivalent.

For large times $P_t(K)$ acquires a LD form \cite{LD}:
\begin{equation}
P_t(K) = {\rm Tr}  \left[ \rho^{(K)}(t) \right] \approx e^{-t \varphi(K/t)} .
\label{phi}
\end{equation}
The function $\varphi(K/t)$ ($k \equiv K/t$) is a ``large-deviation'' function.  It encodes all information about the probability of $K$ at long times \cite{LD}.
An alternative description for the statistics of $K$ is provided by the generating function,
\begin{equation}
Z_t(s) \equiv \sum_{K=0}^{\infty} P_t(K) e^{-s K} .
\label{Zs}
\end{equation}
The ``counting field'' $s$ is the conjugate field to the observable $K$.  In this dynamical context $K$ and $s$ are what  pressure and volume, or magnetization and magnetic field, are in thermodynamical contexts.  Although only the quantum trajectories where $s=0$ are observed in an experiment, we can nevertheless construct a generalized family of quantum trajectories, where the probability of observing $K$ events after time $t$, $P^{(s)}_t(K)$, is biased by the value of $s$ \cite{prl,budini},
\begin{equation}
P^{(s)}_t(K)=\frac{P_t(K){\rm e}^{-sK}}{Z_t(s)}.
\end{equation}
In the long time limit the generating function also acquires a LD form \cite{LD},
\begin{equation}
Z_t(s) \approx e^{t \theta(s)} .
\label{theta}
\end{equation}
The LD functions $\varphi(k)$ and $\theta(s)$ are to trajectories \cite{Lecomte,Garrahan,prl} what entropy density and free-energy density are to configurations in equilibrium statistical mechanics \cite{books}.  The two LD functions are related by a Legendre transform, $\theta(s) = - \min_k \left[ \varphi(k) + k s \right]$ \cite{LD,Mukamel}.  The number of events $K$ is a time-extensive quantity. The scaled activity $k=K/t$  is a {\em dynamical order parameter} as it serves to qualify and distinguish dynamical phases. 

Our approach differs from that often taken with studies of counting statistics\cite{Mukamel} as we focus on exploring the behavior beyond the $s=0$ limit. The function $\theta(s)$ has the convexity properties of (minus) a free-energy.
Furthermore, its analytic properties as a function of $s$ encode non-trivial fluctuation properties of dynamical trajectories.  In particular, singularities in $\theta(s)$ correspond to dynamical (or space-time \cite{Garrahan,prl}) phase transitions.  Just like in equilibrium statistical mechanics, our aim is to compute this free-energy-like function to uncover the dynamical phase structure of the micromaser.

The statistical properties of the $s$-biased trajectories are described by the master equation for $\rho_s(t)=\sum_K\rho^{(K)}(t){\rm e}^{-sK}$,
\begin{equation}
\dot{\rho}_s={\cal W}_s(\rho_s),
\end{equation}
where the generalized quantum master operator (GQMO) reads \cite{Brown,Mukamel}:
\begin{eqnarray}
{\cal W}_s(\rho_s) &\equiv& e^{-s} L_1 \rho_s L_1^\dagger
\nonumber \\
&& + \sum_{\mu \neq \lambda} L_\mu \rho_s L_\mu^\dagger - \frac{1}{2} \sum_{\mu=1}^{N_{\rm L}} \{  L_\mu^\dagger L_\mu , \rho_s \} ,
\label{Ws}
\end{eqnarray}
The LD function $\theta(s)$ is given by the largest eigenvalue of ${\cal W}_s$ \cite{prl} and it is then straightforward to obtain the moments of the distribution $P^{(s)}_t(K)$.

\section{Mean-field analysis}
\label{mfsection}

Although a full analytic solution for the steady-state density operator is readily obtained for $s=0$, this approach is not easily generalized to the case $s\neq 0$. We therefore proceed by using a mean-field approach to capture the behavior of the LD function. We check the accuracy of this calculation using numerical methods in Section \ref{numerics}.

The LD function can be estimated from (\ref{Ws}) through a variational ansatz.  For simplicity we start by considering the limit of zero temperature, $\nu=0$, and we restrict the analysis to density matrices which are diagonal in the number basis.  With the rescalings $a \to a \sqrt{N_{\rm ex}}$ and $a^\dagger \to a^\dagger \sqrt{N_{\rm ex}}$ the GQMO reduces to an operator,
\begin{eqnarray}
\frac{{\cal W}_s}{N_{\rm ex}} \to W_s &\equiv& e^{-s} a^{\dagger } \frac{\sin ^2\left(\alpha  \sqrt{a^{\dagger} a+\delta}\right)}{\sqrt{a^{\dagger} a+\delta}} + \sqrt{a^{\dagger} a+\delta} ~ a
\nonumber \\
&& - \sin ^2\left(\alpha  \sqrt{a^{\dagger}a+ \delta}\right)-a^{\dagger}a ,
\label{Wop}
\end{eqnarray}
where $\delta \equiv N_{\rm ex}^{-1}$.

\begin{figure*}
\includegraphics[width=1.9\columnwidth]{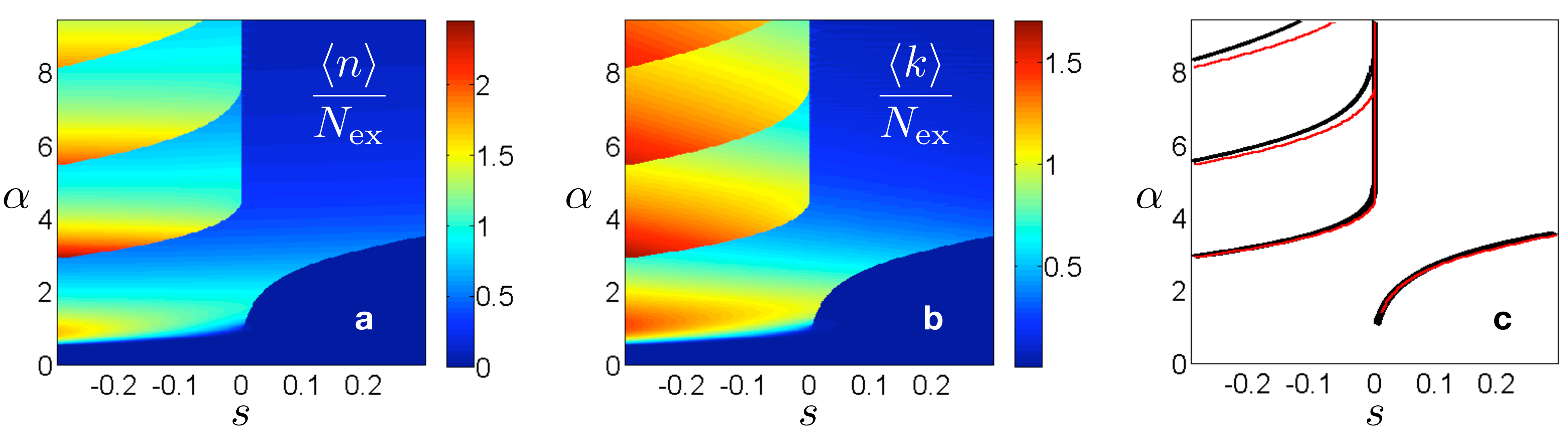}
\caption{(Color online)
(a) Space-time phase diagram from the mean-field approximation ($N_{\rm ex} = 100$ and $\nu=0.15$).  The average cavity occupation $\langle n \rangle / N_{\rm ex}$ is plotted as a function of $\alpha$ and $s$.  (b) The same plot but for the (scaled) activity per unit time, $\langle k \rangle / N_{\rm ex}$.  (c) Comparison of exact (black) and mean-field (red) phase boundaries.
}
\label{fig3}
\end{figure*}

A variational approximation to the largest eigenvalue of $W_s$ is obtained by maximizing $W_s$ w.r.t. $a$ and $a^\dagger$, that is, by maximizing the matrix element of $W_s$ between variational coherent states.  We set $a = e^{i \gamma} \sqrt{n}$ and $a^{\dagger} = e^{-i \gamma} \sqrt{n}$, and the corresponding Euler-Lagrange equations are $\partial W_s / \partial \gamma = 0$ and $\partial W_s / \partial n = 0$.  The first of these can be solved,
\begin{equation}
\frac{\partial W_s}{\partial \gamma} = 0 \Rightarrow \left\{
\begin{array}{lcl}
a &=& e^{-s/2} \sqrt{n} ~ \frac{\left| \sin \left(\alpha  \sqrt{n+\delta }\right) \right|}{\sqrt{n+\delta }} \\
& & \\
a^\dagger &=& e^{s/2} \sqrt{n(n+\delta)} ~ \left|  \csc \left(\alpha  \sqrt{n+\delta }\right)  \right|\\
\end{array}
\right.
\label{mfeq1}
\end{equation}
By replacing these relations into $W_s$ we obtain a variational ``free-energy''
\begin{eqnarray}
{\cal F}_s(n) &\equiv& n - 2 e^{-s/2} \sqrt{n} ~ \left|  \sin \left(\alpha  \sqrt{n+\delta }\right) \right|
\nonumber \\
&& + \sin ^2\left(\alpha  \sqrt{n+\delta }\right) ,
\label{Fmfzero}
\end{eqnarray}
whose minimum w.r.t.\  $n$ gives the variational estimate of the LD function,
\begin{equation}
\theta(s) \approx - \min_n {\cal F}_s(n) .
\label{thetamf}
\end{equation}

A similar procedure can be applied at finite temperature, $\nu \neq 0$.  In this case the variational free-energy reads,
\begin{eqnarray}
{\cal F}_s(n) &=&
(1+\nu)n -2 \sqrt{n (\nu +1)}
\nonumber \\ &&
\times \sqrt{(n+\delta ) \nu + e^{-s} \sin ^2\left(\alpha  \sqrt{n+\delta }\right) }
\nonumber \\ &&
+ \left[ (n+\delta ) \nu + \sin ^2\left(\alpha  \sqrt{n+\delta }\right) \right] .
\label{Fmf}
\end{eqnarray}

The LD function obtained from the variational free energy has multiple singularities. Fig.\ 3(a) shows the average photon occupation of the cavity, $n_*$, that minimizes the variational free energy (\ref{Fmf}), ${\cal F}_s'(n_*)=0$.  The left hand side of the plot, $s<0$, displays multiple transitions where $n_*$ changes discontinuously.  On the right hand side, $s>0$, there is a single first-order line. In contrast to the $s<0$ transitions, this line ends at a critical point, $s_c \gtrsim 0$ and $\alpha_c \gtrsim 1$.  The existence of this critical point is most easily established from the form of ${\cal F}$ at zero temperature, Eq.\ (\ref{Fmfzero}).  In the limit of $\delta =0$ and at $s=0$ this variational free energy reads, $
{\cal F}_s(n) \left( \sqrt{n} - \left|  \sin \left(\alpha  \sqrt{n}\right) \right| \right)^2$.  For $\alpha < 1$ this function is minimized by $n_*=0$; for $\alpha > 1$ it also has a second minimum at $n_* =  \sin^2 \left(\alpha  \sqrt{n}\right)$.  At the critical point $\alpha_c = 1$ these two extrema coalesce.  Furthermore, in the limit of $\delta=0$ the coexistence line is along the $s=0$ axis.  When $\delta \neq 0$ the critical point shifts to $s_c \gtrsim 0$ and $\alpha_c \gtrsim 1$ and the coexistence line bends to the right.

The phase structure revealed by $n_*$ is that of the dynamical phases.   The LD function $\theta(s)$ is the moment generating function for $K$, and $\langle k\rangle(s)=\langle K \rangle(s)/t = - \theta'(s)$.  From Eqs.\ (\ref{thetamf},\ref{Fmf}) we obtain the relation between $n_*$ and $k_* \equiv \langle k \rangle(s) / N_{\rm ex}$ in the mean-field approximation:
\begin{equation}
k_* = \frac{e^{-s/2} \sqrt{n_* (\nu +1)} \sin ^2\left(\alpha  \sqrt{n_* + \delta }\right)}{\sqrt{(n_* + \delta ) \nu + e^{-s} \sin ^2\left(\alpha  \sqrt{n_* + \delta }\right) }} \ .
\end{equation}
The activity $k_*$ has the same phase behavior as $n_*$ as shown in Fig.\ 3(b). The close similarity between the two is easy to understand in terms of the energy balance in the system: at low temperatures the number of quanta absorbed per unit time by the cavity is dominated by the contribution from the atoms, $\langle k\rangle$, whereas the rate of energy loss (to the bath) is proportional to $\langle n\rangle$.

For $\alpha > \pi$ and $s \approx 0$ the phase-diagrams of Fig.\ 3 show that the first-order transition lines converge to $s=0$ with increasing $\alpha$.  The mean-field solution of (\ref{Fmf}) predicts that these transition lines all accumulate at $s=0$ \cite{prl}. 

The mean-field treatment of this section will become less accurate for larger values of $\alpha$ where non-linearities in the operator (\ref{Wop}) become more prominent.  The approximation above assumes, firstly, that averages of products of $a$ and $a^\dagger$ can be reduced to products of their averages, and, secondly, that the operator (\ref{Wop}) is normal ordered.  Furthermore, the approximation assumes that the eigenstate associated to the maximal eigenvalue $\theta(s)$ is a coherent state.  This means that when $\alpha \gg 1$ the mean-field treatment may not give accurate estimates of $\langle n \rangle$, for example near the dynamical crossovers at $\alpha \approx 2 \pi, 4 \pi, \ldots$ at $s=0$ [see Fig.\ 2(b)].  Nevertheless, the dynamical Landau free-energy (\ref{Fmf}) does capture the singular changes in $\theta(s)$, so we expect the mean-field treatment here to give a good approximation to the dynamical phase-behaviour of the micromaser, as is confirmed in the next section where we compare to the exact numerical diagonalisation of the GQMO.

\section{Exact numerical diagonalisation of the GQMO}
\label{numerics}

For the numerical derivation of the LD function we constructed the matrix representation of the GQMO for given $N_{\rm ex}$ and $\nu$ in the number state basis.  The matrix was truncated at a state with occupation number $N_{\rm max}$ significantly larger than $N_{\rm ex}$ and the largest eigenvalue was obtained as a function of $s$ and the pump parameter $\alpha$. The activity $K$ and other quantities were calculated by taking numerical derivatives of the LD function.

\begin{figure}
\includegraphics[width=.9\columnwidth]{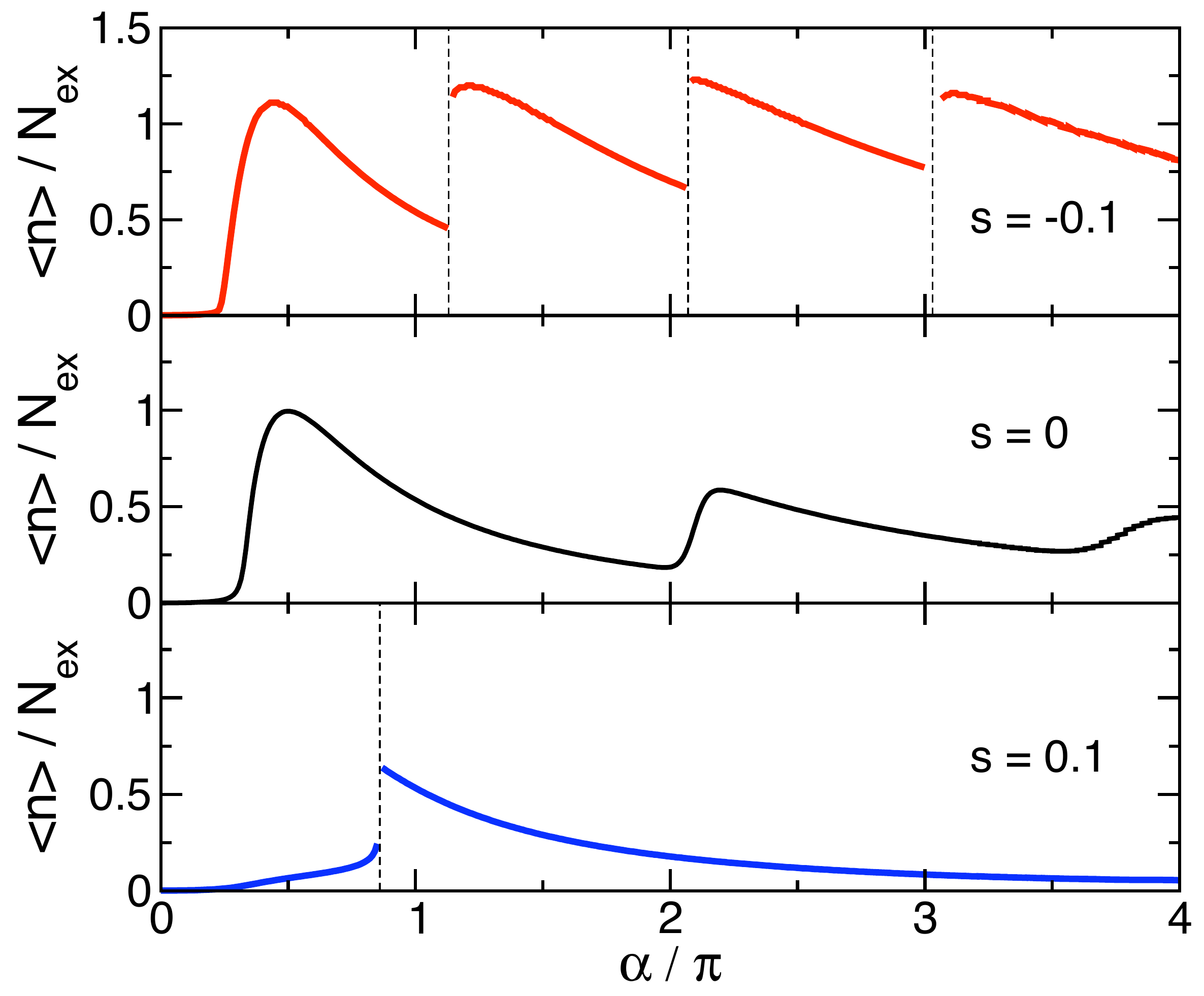}
\caption{(Color online)
Average cavity occupation, $n$ (scaled by $N_{\rm ex}$), as a function of pump parameter, $\alpha$, for different values of the counting field $s$ (see Figs.\ 1 and 3 for reference).  (Top) Active side of the dynamics ($s < 0$): $n$ is discontinuous at the first-order transitions (their location indicated by dashed lines).  
(Bottom) Inactive side ($s > 0$): here $s > s_c \approx .0087$, and there is a single first-order jump in $n$.  (Center) For comparison, the same curve at normal dynamical conditions, $s=0$, displays no singularities.
}
\label{fig4}
\end{figure}

The resulting phase diagram is shown in Fig.\ \ref{fig1}.  There we show the average activity as a function of pump parameter $\alpha$ and counting field $s$.  The dynamical or space-time phase behavior is very similar to that obtained from the mean-field approximation.  Fig.\ \ref{fig3}(c) shows a comparison of the exact and mean-field phase boundaries.  These are very close to each other, and only start to disagree at large $\alpha$: this is to be expected since the larger $\alpha$ the more prevalent non-linearities become which are captured less accurately by the mean-field expansion.

Figure \ref{fig4} shows the scaled average cavity occupation, $\langle n \rangle / N_{\rm ex}$, as a function of pump parameter for different $s$.  At $s=0$ the curve is smooth;  this reproduces the plot shown in Fig.\ \ref{fig2} (b): the photon occupation oscillates with varying pumping, in a way which is suggestive of phase transitions without actually being singular at finite $N_{\rm ex}$. In contrast, for $s\neq0$ we find discontinuous behavior even for the relatively small value of $N_{ex}=100$ chosen.

For $s < 0$ the curves display singularities at the values of $\alpha$ where the transition lines are crossed, see Fig.\ \ref{fig3}.  At these points the average occupation and the dynamical activity jump by a finite amount, so they are first-order transitions. These first-order phase boundaries  do not seem to end, see Fig.\ 1, so $\langle n\rangle$ and $\langle k\rangle$ show qualitatively the same behavior for all negative $s$.  The sizes of the jumps $\langle n\rangle$ at the first-order transitions get smaller at larger $\alpha$ which is reflected in the behavior at $s=0$ where the crossovers get more and more smeared out as $\alpha$ is increased.  For $s$ positive and larger than the critical value $s_c$, the occupation vs. pumping curve shows one singularity when the rightmost transition line is crossed.

\begin{figure}
\includegraphics[width=\columnwidth]{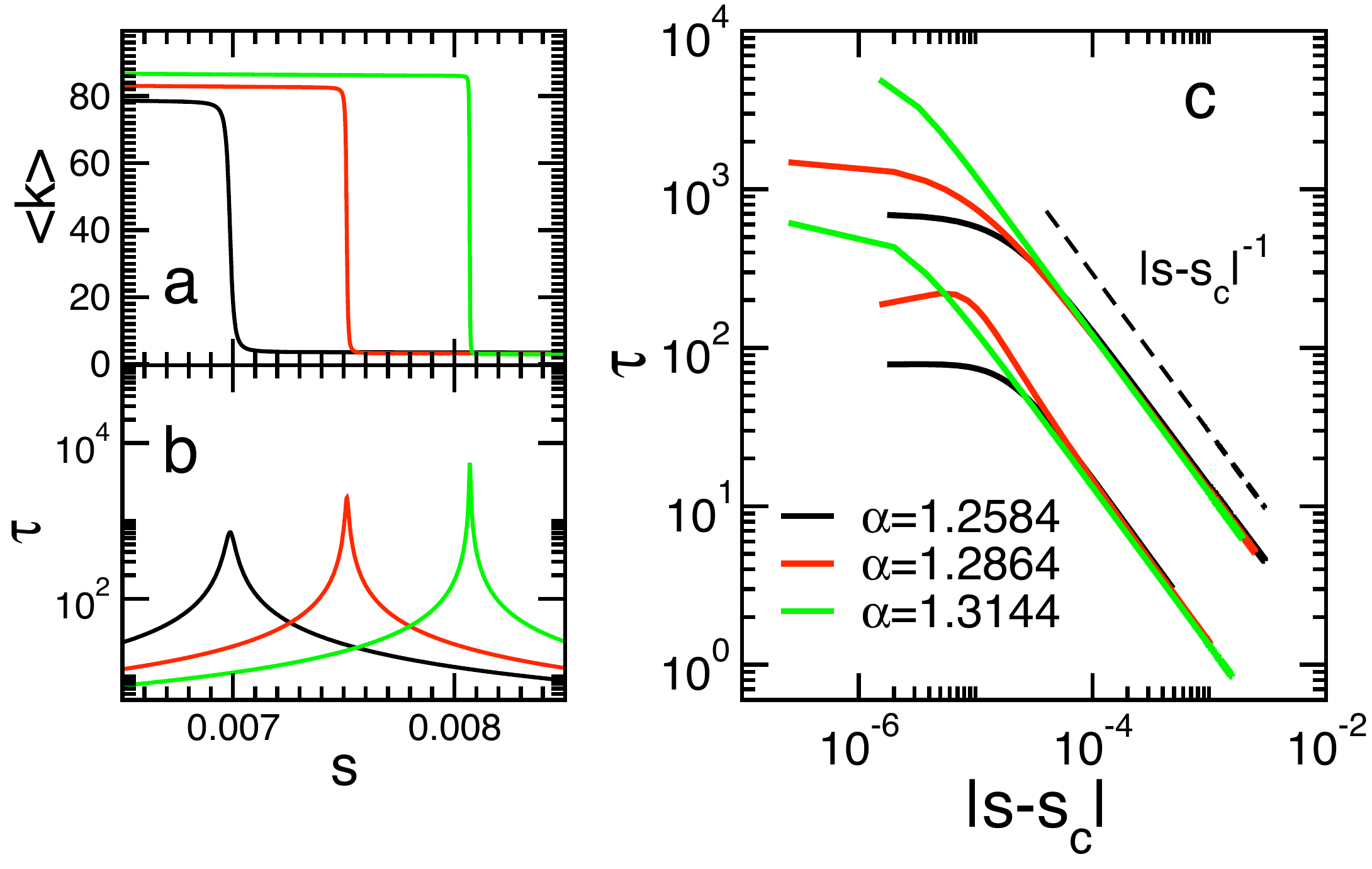}
\caption{(Color online) 
Diverging correlation time at the critical point.  (a) Average dynamical activity as a function of $s$ for three values of the pump parameter $\alpha$ approaching the critical value $\alpha_c \approx 1.342$.   (b) Correlation time $\tau$ obtained from the difference between the leading and subleading eigenvalues of the GQMO.  (c) The correlation time appears to diverge at $(s_c, \alpha_c)$ as $\tau \sim |s-s_c|^{-1}$, approaching both from the left and right (the left branch has been shifted down by a constant factor for clarity).  Results from numerically exact diagonalisation of the GQMO, Eq. (\ref{Ws}); the Hilbert space was truncated at $N_{\rm max} = 350$, where $N_{\rm ex}=100$.
}
\label{fig5}
\end{figure}

The critical point is considered in more detail in Fig.\ \ref{fig5}.  Panel (a) shows the dynamical activity as a function of $s$ for subcritical values of $\alpha$ approaching $\alpha_c \approx 1.342$.  Panel (b) shows the corresponding correlation time $\tau$ for quantum trajectories.  This is calculated from the gap between the leading eigenvalue of the GQMO, i.e. the LD function $\theta$, and the subleading one $\theta_2$: $\tau \equiv (\theta - \theta_2)^{-1}$.  This is just like calculating the correlation length in a thermodynamic problem where the partition sum can be obtained from a transfer matrix \cite{books}.
The correlation time appears to diverge at the critical point as $\tau \sim | s - s_c |^{-1}$, see Fig.\ \ref{fig5}(c).

\section{Conclusions and Discussion}

The approach we have taken in this paper offers a new perspective on the dynamical transitions which occur in open quantum systems such as the micromaser. Instead of focusing on the steady state solution of the master equation, we have examined the statistical properties of the quantum trajectories corresponding to the measurement of the atoms after they leave the cavity. We used the large deviation limit as a way of exploring the analytic properties of the moment generating function at non-zero values of the counting field, $s$. Surprisingly, we were able to identify discontinuities in the large-deviation function for non-zero values of $s$ which we interpret as corresponding to first order phase transitions. We also found a critical point where a first order transition line terminates and which we found displays the standard scaling behavior associated with a continuous thermodynamic phase transition.

The non-linear character of the quantum master equation (\ref{QME}) gives rise both to the phase transitions we find at $s\neq 0$ and the well-known dynamical crossovers which are found from an analysis of the steady-state properties of the micromaser.  For $\alpha\simeq 1$, one can think of the $s=0$ behavior as being controlled by the nearby critical point at $(\alpha_c,s_c)$. Furthermore, we showed that the critical point moves towards the $s=0$ line as $N_{ex}$ is increased, reaching it in the limit $N_{ex}\rightarrow \infty$, which is entirely consistent with the interpretation of the dynamical crossovers found from the steady-state density operator as true phase transitions in this limit.  The proximity of the first-order transition lines to $s=0$ at higher values of $\alpha$ will also have observable consequences in terms of dynamical fluctuations (think for example of the prominence of vapor bubbles in a liquid at conditions near liquid-vapor coexistence).

Even though it is just the unbiased trajectories that can be measured in an experiment, our predictions about the presence of singular behavior at non-zero $s$ could still be tested quite easily as the statistics for all the biased trajectories can be generated when those for the unbiased case are known. However, a detailed comparison with experimental data would require us to extend our model to include the imperfections that typically occur in real measurements of the atomic state. It may also require the issue of finite-time corrections to be considered. 

In this paper we have restricted ourselves to studying the simplest model of a micromaser and to examining only the most basic statistical properties of the trajectories. It will be interesting to examine more complex models where quantum coherences play an important role and the dynamics is not necessarily Markovian, as well as examining the dependence on $s$ of higher cumulants of the counting distribution. We expect that the real utility of the method will be in analyzing more complex systems where a full understanding of the steady-state density matrix is not easily obtained.

\acknowledgements
This work was supported in part by EPSRC grant no.\ EP/H024069/1. 

\end{document}